\documentclass[preprint2,6pt]{emulateapj}%
\usepackage{graphicx}
\usepackage{amsmath}
\usepackage{amsfonts}
\usepackage{amssymb}%
\providecommand{\U}[1]{\protect\rule{.1in}{.1in}}

\begin{document}

\title{Monte Carlo simulations of the photospheric emission in GRBs}
\shorttitle{MC simulation of photospheric emission}
\author{D. B\'egu\'e \altaffilmark{1,2,3}, I. A. Siutsou\altaffilmark{1,2}, G. V. Vereshchagin\altaffilmark{1,2}}
\shortauthors{B\'egu\'e et al.}

\submitted{Accepted to ApJ}

\altaffiltext{1}{University of Roma ``Sapienza'', 00185, p.le A. Moro 5, Rome, Italy}
\altaffiltext{2}{ICRANet, 65122, p.le della Repubblica, 10, Pescara Italy}
\altaffiltext{3}{Erasmus Mundus Joint Doctorate IRAP PhD student}

\begin{abstract}
We studied the decoupling of photons from ultra-relativistic spherically symmetric outflows
expanding with constant velocity by means of Monte-Carlo (MC) simulation.
For outflows with finite width we confirm the existence of
two regimes: photon thick and photon thin introduced recently by Ruffini, Siutsou, Vereshchagin (2011), hereafter RSV.
The probability density function of photon last scattering is shown to be very
different in these two cases. We also obtained spectra as well as light curves. In photon thick case, the time
integrated spectrum is much broader than the Planck function and its shape is well
described by the fuzzy photosphere approximation introduced by \cite{RSV}. In the photon
thin case we confirm the crucial role of photon diffusion, hence the probability density of decoupling has a maximum
near the diffusion radius, well below the photosphere. Its spectrum has Band shape. It is produced when the outflow is optically thick and its peak is formed at diffusion radius.
\end{abstract}
\maketitle

\section{Introduction}

Photospheric emission is considered among the leading mechanisms generating
gamma rays from relativistically expanding plasma in cosmic Gamma Ray Bursts
(GRBs) \citep{goodman}. In both fireball \citep{piran99} and fireshell \citep{ruffini} models photospheric emission
is naturally produced when expanding initially optically thick plasma becomes
transparent for photons.

While within the fireball model the basic approximation involved is that of
(infinitely long) steady {\em wind}, in the fireshell model the geometry of the outflow
is a thin {\em shell}. It is shown by \cite{RSV} that in both wind and shell models
when outflows with {\em finite} duration are considered some important aspects of the
optical depth behaviour are overlooked. There a new physically motivated classification was introduced:\ photon thick
and photon thin outflows with respectively $R_{ph}<R_t$ and $R_{ph}>R_t$ with $R_{ph}$ being the
photospheric radius, $R_t=2\Gamma^2 l$ the radius of transition, $\Gamma$ is the Lorentz factor and $l$
the laboratory width of the outflow. Light curves and
spectra from finite outflows were obtained using two different
approximations: sharp and fuzzy photosphere. These approximations were derived
there from the transport equation for radiation.

In this work we examine the validity of these approximations using MC simulations of photon
scattering. We compute the probability
of last scattering in finite relativistic outflows. Then we obtain light
curves and spectra from both photon thick and photon thin outflows, and compare them with the results
obtained before.

\section{The model}

Following \cite{RSV} we consider a spherical wind of finite width $l$ with laboratory
number density of electrons being zero everywhere except the region
$R(t)<r<R(t)+l$ where it is

\begin{equation}
n_{l}=n_{0}\left(  \frac{R_{0}}{r}\right)  ^{2}\label{portionofwind}, ~ ~ ~ \Gamma(r)=\text{const},
\end{equation}
with $R_{0}$ and $n_{0}$ being respectively the radius and the density at the base of the outflow. This outflow expands with constant Lorentz factor $\Gamma$, so that $R(t)=R_{0}+\beta ct$, where $\beta=\sqrt{1-\Gamma^{-2}}$. Accelerating outflows are not considered here. Notice that when there is strong relative difference of Lorentz factors within the outflow with $\Delta\Gamma\sim \Gamma$ a radial spreading phenomenon discussed in \cite{piran93} and \cite{meszaros93} occurs. The outflow width starts to increase linearly with $R(t)$ at the radius
\begin{align}
R_s=\Gamma^2 l \frac{\Gamma}{\Delta\Gamma}.
\end{align}
It is clear from this formula that such phenomenon occurs at radii much larger than $\Gamma^2 l$ when $\Delta\Gamma\ll \Gamma$. In what follows we consider such case only. Our results are valid for arbitrary constant $l$. The expression for the optical depth in the model defined by Eq. (1) is found in \cite{RSV} and it reads:

\begin{align}
\tau( &  r,\theta,t)=\tau_{0}R_{0}\left[  \frac{\theta-\tan^{-1}\left(
\frac{r\sin\theta}{ct+r\cos\theta}\right)  }{r\sin(\theta)}\right.
\label{tau}\\
&  \left.  -\beta\left(  \frac{1}{r}-\frac{1}{\sqrt{(ct+r\cos\theta
)^{2}+(r\sin\theta)^{2}}}\right)  \right]  ,\nonumber
\end{align}
where%
\begin{equation}
\tau_{0}=\sigma n_{0}R_{0}=\frac{\sigma E_{0}}{4\pi m_{p}c^{2}R_{0} l\Gamma
},\label{tau0}%
\end{equation}
and $r$ is the position of photon emission, $\theta$ is the angle between
momentum of photon and the radius vector, $\sigma$ is the Thomson cross
section, $t$ is the time the photon remains within the outflow, $E_{0}$ is
total energy release in the GRB, $m_{p}$\ is the proton mass, and $c$\ the speed of
light. Time $t$ is found via the equations of motion of the outflow and
that of the photon.

The optical depth in finite relativistic outflows has two different asymptotic, depending
on initial conditions: in photon thick case $\tau$ is almost constant within the
outflow; instead in the photon thin case it is linearly increasing with
depth from the outer boundary at $R(t)+l$.

In \cite{RSV} diffusion of photons within the outflow is found to play a crucial
role in photon thin case: the radius at which photons effectively diffuse out
of the outflow is smaller than the photospheric radius $R_{ph}$. The latter is
defined by equating expression (\ref{tau}) to unity for emission with $\theta=0$
and $t$ corresponding to the inner boundary. Hence, when the photon thin outflow
reaches the photospheric radius there are few photons left in it.

\section{Monte Carlo simulation}

We have used a spherically symmetric MC simulations of
photon scattering inside the outflow described by Eq. (\ref{portionofwind}). For more detail on interactions
in relativistic expanding plasma (see B\'egu\'e, Siutsou and Vereshchagin, in preparation). In this
simulation each photon is followed as it interacts with electrons until it
decouples from the outflow. Photons are injected in the outflow when the optical depth of the inner boundary is $\tau_i$ given in Tab. \ref{Tab:para}.
It is also assumed that their initial distribution is isotropic and
thermal. The radial position of each photon inside the outflow is chosen randomly.

The code consists mainly of a loop computing each scattering.
We proceed in two steps. First an infinite and steady wind (already
treated by \cite{Beloborodov11} with $l\rightarrow\infty$ and $R\rightarrow 0$) is considered. For a given
position characterized by $r$ and $\theta$ of photon in the outflow, we
compute a maximal value for the optical depth $\tau_{max}$ using (\ref{tau}) with
$t\rightarrow\infty$. The probability for the photon to decouple the outflow is
$\exp\left(  -\tau_{max}\right)  $. Then a random number $X\in\lbrack0,1]$ is
chosen. On the one hand, if $X<\exp\left(  -\tau_{max}\right)  $ the photon is considered decoupled.
Afterwards photon remains in the outflow but does not
scatter. This case corresponds to decoupling in photon thick case,
for which the presence of boundaries is not essential. On the other hand, if
$X\geq\exp\left(  -\tau_{max}\right)  $ the position $r_{s}$ of next
scattering is computed from the optical depth%
\begin{equation}
\tau(r,r_{s},\theta)=\tau_{0}R_{0}\left[  \frac{\theta-\theta_{s}}%
{r\sin(\theta)}-\beta\left(  \frac{1}{r}-\frac{1}{r_{s}}\right)  \right]
\end{equation}
where $\theta_{s}$ is the angle between photon momentum and the radius vector
at the position of scattering.\ The new radial position $r_{s}$ such that
$\exp\left[  -\tau(r,r_{s},\theta)\right]  \equiv X$ is found by iterations.

The second step is to take into account the finite width of the outflow. To do so,
$r_{s}$ is compared with the radii of the inner and outer boundaries of the
outflow. If $r_{s}<R(t_s)$ or $r_{s}>R(t_s)+l$ the scattering does not take place
and the photon is considered decoupled. Such decoupling occurring at the boundaries
corresponds to the photon thin case. In the opposite case
$R(t_s)\leq r_{s}\leq R(t_s)+l$ the scattering is assumed to occur and the loop is repeated until the photon decouples.

We consider two models of scattering: coherent
isotropic scattering and Compton scattering. The former is treated for
comparison with results in the literature such as e.g. \cite{Beloborodov11}. Such model is also
interesting {\it per se} since it preserves Planck spectrum and it traces only geometrical
effects. Instead when Compton scattering is considered, the equilibrium spectrum
is the Wien one because stimulated emission is not taken into account by MC simulation. For initially large optical depth the spectrum indeed firstly acquires the Wien shape, and only at the photosphere it changes the shape again.

The coherent scattering is computed by Lorentz transformation to the reference
frame comoving with the outflow. In addition, when Compton scattering is
considered, another Lorentz transformation to the rest frame of the electron is performed.
The electron is chosen randomly from the Boltzmann distribution
at a given comoving temperature defined as%
\begin{equation}
T=T_{0}\Gamma^{-1/3}\left(  \frac{R_{0}}{r}\right)  ^{2/3},\qquad
T_{0}\simeq\left(  \dfrac{3E_{0}}{4\pi aR_{0}^{3}}\right)  ^{1/4},
\label{EqT}
\end{equation}
where $a=4\sigma_{SB}/c$, $\sigma_{SB}$\ is the Stefan-Boltzmann constant,
see e.g. \cite{RV}.

Various tests of the code are performed. Before dealing with finite outflows we
also reproduced the results of \cite{Beloborodov11}. We present below the results for two specific cases.
The set of parameters for these simulations are given in Tab. \ref{Tab:para}. The left (right) column corresponds to photon thick
(thin) case.

\begin{table}[ht!]
\caption{Parameters for the simulations.}%
\label{Tab:para}%
\centering%
\begin{tabular}
[c]{|c|c|c|}\hline
& Photon thick & Photon thin\\\hline
$R_{0}$ (cm) & $10^{8}$ & $10^{8}$\\\hline
$l$ (cm) & $4.5\ast10^{8}$ & $10^{8}$\\\hline
$\tau_0$ & $2.3\ast10^{10}$ & $1.2\ast10^{13}$\\\hline
$E_{0}$ (erg) & $1.5\ast10^{52}$ & $10^{54}$\\\hline
$\Gamma$ & $500$ & $300$\\\hline
$R_{ph}$ (cm) & $4.6\ast10^{12}$ & $3.3\ast10^{14}$\\\hline
$R_{D}$ (cm) & $3.5\ast10^{13}$ & $1.0\ast10^{14}$\\\hline
$\tau_i$ & $10^2$ & $1.3\ast10^3$\\\hline
\end{tabular}
\end{table}

\section{Results}

\subsection{Probability density function of position of last scattering}

\begin{figure}[ht!] \centering
\begin{tabular}[c]{c}%
\includegraphics[scale=0.60]{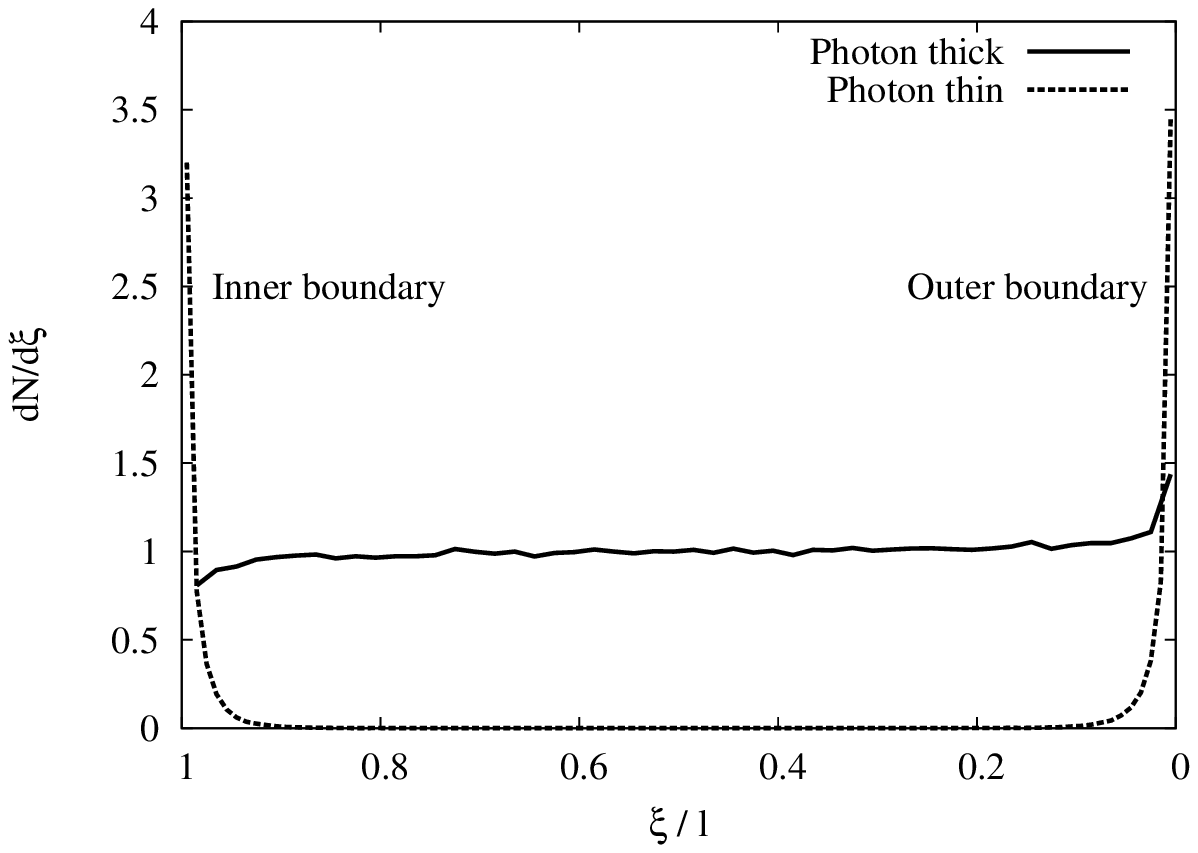}\\
\includegraphics[scale=0.60]{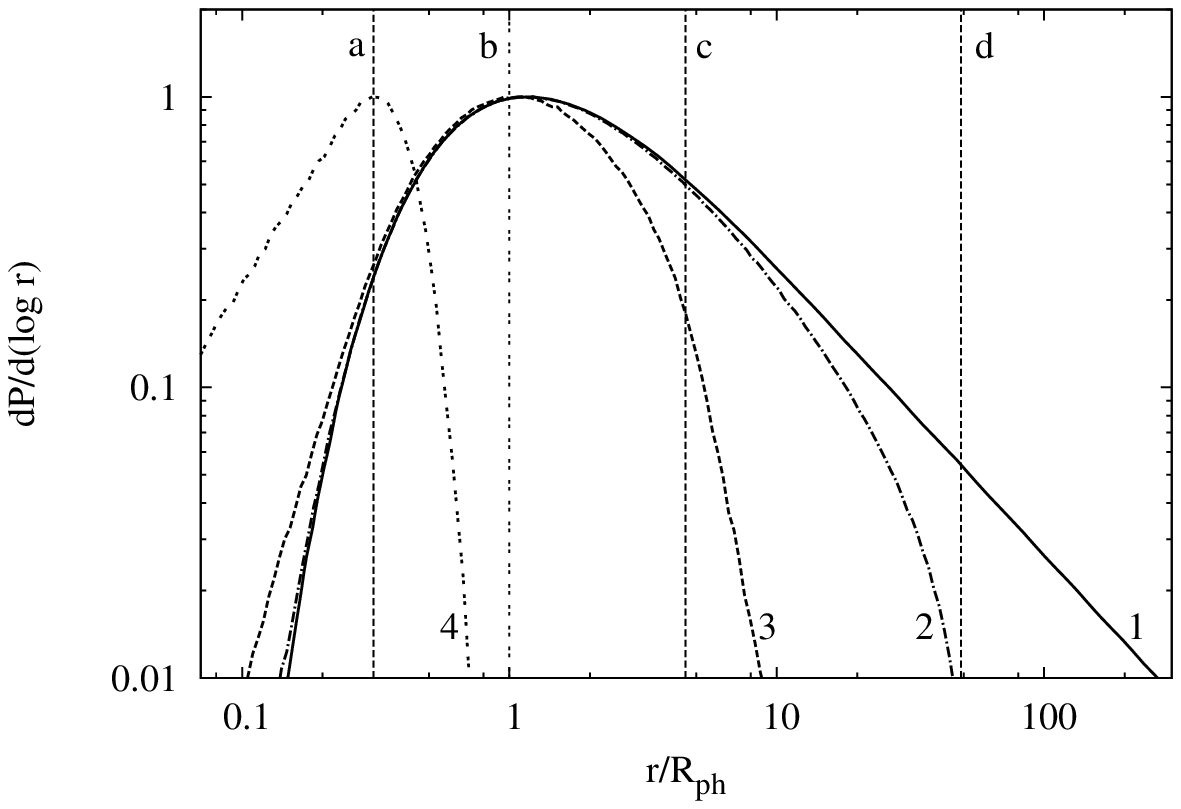}
\end{tabular}
\caption{Upper panel: Probability density of last scattering
as a function of normalized depth for photon thick and photon thin
outflows (in the latter case decreased by a factor of 10 for a better presentation). \\
Lower panel: probability density function for the position of last
scattering in the following cases: infinite and steady wind (1), photon thick case of Tab. \ref{Tab:para} (2), photon thick case of Tab. \ref{Tab:para} with a different Lorentz factor $\Gamma=300$ (3), and photon thin outflow (4).
The vertical line (a) represents the diffusion radius in photon thin case, line (b) represents the photospheric radius while lines (c) and (d) show the transition radius $R_t$ for curves (2) and (3), respectively. Both coherent and Compton scattering models give the same results.}%
\label{Fig:proba}%
\end{figure}

Consider first the probability density of photon last scattering position as a function of
depth (top panel of Fig. \ref{Fig:proba}). Photon decoupling from
photon thick outflows is expected to be local and the presence of boundaries should not change
this probability substantially, indeed the probability distribution function of
last scattering is found to be almost independent on the depth.
On the contrary in photon thin outflow there is enough time for photons to be transported
at the boundaries by diffusion as discussed in \cite{RSV}. As a result the probability density is peaked at the boundaries.

Difference between photon thin and photon thick outflows is also reflected in
the probability density of last scattering as a function of the radius, shown
at the bottom of Fig. \ref{Fig:proba}. In the photon thin outflow most photons
escape from the outflow well before the photospheric radius, namely near
diffusion radius
\begin{equation}
R_{D}=\left(  \tau_{0}\Gamma^{2}R_{0}l^2\right)  ^{1/3}.
\end{equation}

Instead for photon thick outflows the probability density function of last
scattering is found to be close to the one of infinite and steady wind, found
by \cite{Beloborodov11}. The finite extension of the outflow results in the
exponential cut-off for the probability density function at radii larger than $R_t$.

Let us stress how different the probability densities for photon thick and photon
thin cases are. Not only the positions of the maximum, but also the shapes are
different. In photon thin case at small radii the number of photons that
diffuse out is determined by the change in the diffusion coefficient, i.e.
the electron density which follows a power law. In expanding plasma while the density decreases,
the mean free path for photons increases which makes the diffusion faster. At
large radii almost all photons have already diffused out, and the probability has an
exponential cut off. As for the photon thick case, the probability density dependence
on the radius is opposite. For small radii the probability for last scattering
is exponentially suppressed because of large optical depth, while at large radii
it follows a power law, as discussed in \cite{peer2008}. At larger radii we find an exponential
decrease due to the finiteness of the outflow, see Fig. \ref{Fig:proba}.

\begin{figure}[t]
\centering
\includegraphics[scale=0.90]{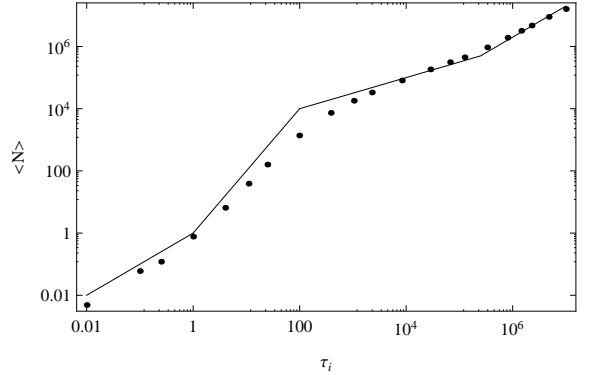}
\caption{Average number of scatterings as function of initial optical depth $\tau_i$. Results of MC simulations are shown with points, while analytic results in different asymptotic from Appendix are shown by the corresponding lines. In the optically thin regime the number of scatterings is proportional to $\tau_i$. For $1<\tau_i<\tau(R_D)$ it is proportional to $\tau_i^2$. For $\tau(R_D)<\tau_i<\tau(R_t)$ it is proportional to $\tau_i^{1/2}$. For even larger $\tau_i$ corresponding to the photon thick asymptotic the number of scatterings is again proportional to $\tau_i$. In these simulations we used $\Gamma=100$, $l=10^8$ cm and $\tau_0=10^{14}$.}%
\label{Fig:num_scatt}%
\end{figure}
We also found the difference between photon thick and photon thin cases when computed the average number of scatterings as function of initial optical depth, shown in Fig. \ref{Fig:num_scatt}. Derivation of analytic results is reported in Appendix.

\subsection{Spectra and light-curves from photon thick outflow}

\begin{figure}[ht] \centering
\includegraphics[scale=0.60]{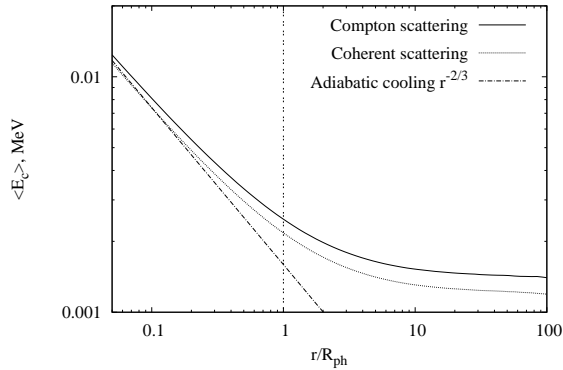}
\caption{Average comoving energy of photons in photon thick outflow as function of the radius. The averaging is performed for photons still coupled with the outflow.}%
\label{Fig:energycom}%
\end{figure}

The average comoving energy of photons for steady wind was considered by \cite{peer2008} and \cite{Beloborodov10}.
For finite outflow we show this quantity in Fig. \ref{Fig:energycom} and compare it with the $r^{-2/3}$
dependence characteristic of adiabatic cooling. The difference between Compton and coherent scattering comes from the fact that the 
average energy in the optically thick regime is $3kT$ for Wien spectrum and $2.82kT$ for Planck spectrum.
In agreement with previous studies we find that the average energy of photons
at the photosphere is higher than adiabatic cooling predicts. It is $1.58$ in the case of Compton scattering
and $1.4$ in the case of coherent one. 

\begin{figure}[ht] \centering
\includegraphics[scale=0.60]{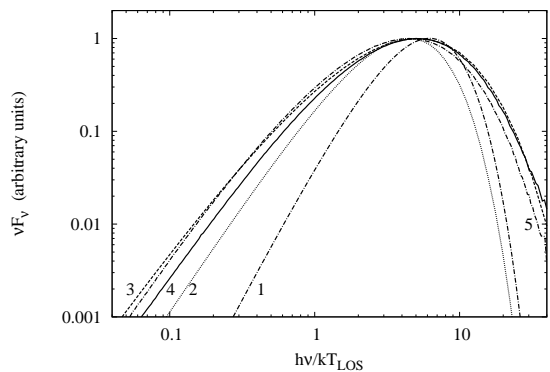}
\caption{
Time integrated spectrum of photospheric emission from photon thick outflow computed within different approximations:
Planck spectrum (1),
sharp photosphere approximation (2),
fuzzy photosphere approximation (3),
Monte Carlo simulation for Compton scattering (4),
Monte Carlo simulation for coherent scattering (5).
Spectra (1), (2) and (3) are shifted in energy by a factor 1.58 (see text for details).
$T_{LOS}$ is the laboratory temperature of the outflow at the photospheric radius.}%
\label{Fig:thickspec}%
\end{figure}

In Fig. \ref{Fig:thickspec} we compare time integrated spectra of photon thick outflow obtained with different models.
Each of them involves $5\ast10^7$ photons.
In order to compare the results of \cite{RSV} with our MC simulations we shifted the spectra for sharp,
fuzzy approximations and Plank spectrum by the factor 1.58, which takes into account the effect discussed above.
The results obtained in fuzzy approximation are in good agreement with MC simulations. The later gives the low energy spectra index 
$\alpha=0.24$ for Compton scattering and $\alpha=0.19$ for coherent one to be compared with $\alpha=0$ found in \cite{RSV}.

\begin{figure}[h] \centering
\includegraphics[scale=0.60]{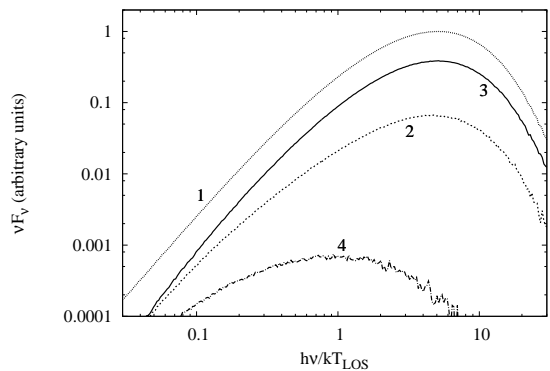}
\caption{Time integrated spectrum from photon thick outflow (1) together with time resolved spectra corresponding to the
beginning of the plateau in the light curve (2), middle of the plateau
(3) and the tail (4).}%
\label{Fig:instant}%
\end{figure}

We also show in Fig. \ref{Fig:instant} time resolved spectra at
selected arrival times which are clearly far from being
Planckian. The effect of spectral flattening at late time discovered by
\cite{Peer} is visible.

\begin{figure}[h] \centering
\includegraphics[scale=0.60]{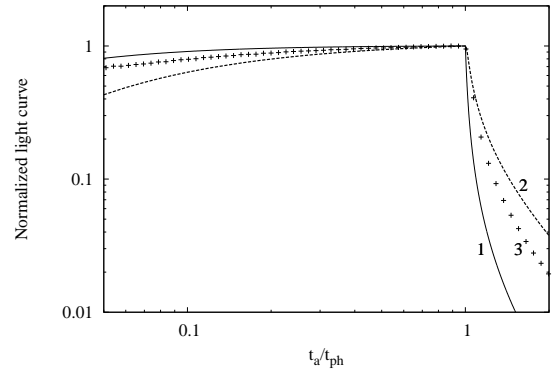}
\caption{Light curves from photon thick outflow within different approximations: sharp photosphere (1), fuzzy photosphere (2) and MC simulation (3) for Compton scattering model.
Coherent scattering model gives similar results.}%
\label{Fig:lc}%
\end{figure}

Finally, light curves are presented in Fig. \ref{Fig:lc}, where we defined $t_{ph}=l/c $ and $t_a=t_e-\mu R_e/c$, with $r_e$
and $t_e$ laboratory radius and time of emission. We chose $t_a=0$ for a photon emitted at $R_e=R_0$. The results of MC simulation are in qualitative agreement with both approximations, slightly favouring the sharp photosphere one.

~

\subsection{Spectra and light curves from photon thin outflow}

In photon thin case diffusion has been shown to be important. Both light curve and spectrum are characterized by the radius of diffusion $R_D$ and the associated arrival time $t_D=R_D/(2 \Gamma^2c)$. It should be noted that the  photospheric emission carries only a fraction of the total energy of a GRB. In the photon thin case the duration of the photospheric emission is relatively short, see the formula above, which implies that the luminosity of this component can be comparable to the luminosity of the rest of the prompt emission \citep{bernadini07}. Such short and intense photospheric emission preceding the main burst has been reported in many GRBs \citep{ryde09}.

Since photons are still strongly coupled to the outflow, local thermodynamic equilibrium is maintained everywhere but at the very boundary. Indeed, this part of the outflow looses photons by random walks. For MC simulations of the photon thin outflows we used the coherent scattering model. The results are compared with \cite{RSV} where light curve and spectrum are obtained using the sharp photosphere approximation, after solving radiative diffusion equation. It has been demonstrated there that the temperature at the boundary of the outflow follows a different law from the one of adiabatic cooling: $T(r) \varpropto r^{-13/24}$ with both temperatures nearly coinciding at the radius of diffusion. So we expect the time integrated spectrum obtained by MC simulations has more power at energies above the peak.

Finally, in both MC simulations and the work of \cite{RSV} the photons emitted outside the relativistic beaming cone from the photon thin outflows are neglected. This is good approximation, since such photons have much lower energy due to the Doppler effect. Besides, their arrival time is $2\Gamma^2$ longer than the arrival time of diffusion.

\begin{figure} \centering
\includegraphics[scale=0.60]{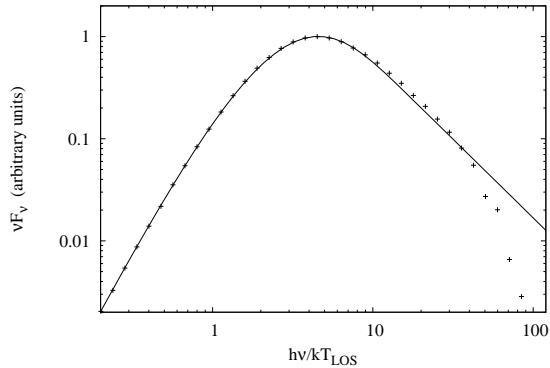}
\caption{Time integrated spectrum from photon thin outflow obtained by MC simulation (crosses), and sharp photosphere approximation (thick curve). $T_{LOS}$ is here the laboratory temperature at $R_D$.}%
\label{Fig:specthincase}%
\end{figure}

\begin{figure} \centering
\includegraphics[scale=0.60]{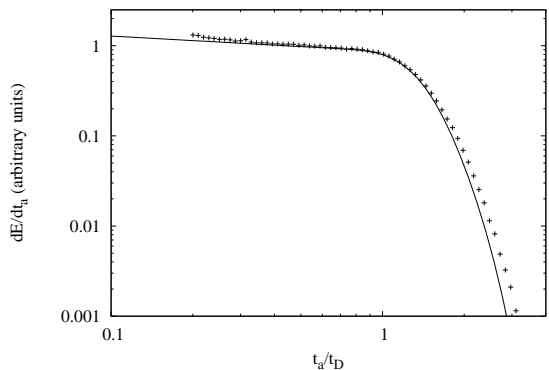}
\caption{Light curves from photon thin outflow obtained by MC simulation (crosses) and sharp photosphere approximation (thick curve).}%
\label{Fig:lcthin}%
\end{figure}

Spectra and light curves are shown respectively in Figs. \ref{Fig:specthincase} and \ref{Fig:lcthin}. The simulation involves $10^6$ photons. We see a good agreement between both methods of computation. The high energy tail with a power law in the time integrated spectrum is clearly visible. Our MC simulation shows the exponential cut off at the energy corresponding to the temperature of photons at the radius defined by the initial optical depth $\tau_i$ (see Fig. \ref{Fig:specthincase}). The power law above the peak energy should extend up to initial temperature $T_0$, see \cite{RSV}. This also results in the missing initial part of the light-curve (see Fig. \ref{Fig:lcthin}).

The power law index is $\beta=-3.5$ for the model of the outflow given by Eq. (\ref{portionofwind}). Such power law is steeper than typical high energy power laws observed in GRBs. It remains to be analyzed how $\beta$ changes for different hydrodynamic models of the outflow. Besides, high energy GeV emission cannot be explained with this model since there exists an exponential cut off corresponding to the initial temperature in the source of GRB, typically being in MeV region.

\section{Conclusions}

To summarize, in this work we considered the photospheric emission from finite relativistic outflow by MC simulations of photon scattering.
The validity of the assumptions made in
\cite{RSV} is verified and it is found that the fuzzy photosphere approximation
used in that paper is adequate in describing both the light curves and the shape of the spectra of
photospheric emission from photon thick outflows. The sharp photosphere approximation produces qualitatively acceptable results for photon thick outflows, but underestimates high energy part of the spectrum and the decaying part of the light curve.

As expected, diffusion is found to dominate the photospheric emission of
photon thin outflows. The probability density of last scattering is obtained
in this case both as function of outflow depth and its radial position. Both
these functions differ from their photon thick counterparts, once more pointing out the essential
difference between these two cases. Good agreement is obtained between MC simulation and sharp photosphere approximation for both light curve and spectrum from photon thin outflow.

\acknowledgments DB is supported by the Erasmus Mundus Joint Doctorate Program
by Grant Number 2011-1640 from the EACEA of the European Commission.

\appendix

\section{Average number of scatterings}

We derive here the average number of scatterings expected for photons in both photon thin
and photon thick outflows.

From \cite{RSV} we recall the expressions of the optical depth for these two cases:
\begin{equation}
   \tau= \left \{
         \begin{aligned}
            & \frac {\tau_0}{2\Gamma^2} \left (  \frac {R_0}{r}  \right ) , & \Gamma R_0 \ll r \ll \Gamma^2 l \label{Eq:tauthick}\\
            & \tau_0 \frac {R_0 l}{r^2} , & r \gg \Gamma^2 l
         \end{aligned}
  \right.
\end{equation}

The average number of scatterings is defined as
\begin{equation}\label{int}
\langle N \rangle=\int_{t_i}^{t_f} \frac{c dt_c}{\lambda_c}=
\int_{t_i}^{t_f} \sigma \langle n_c\rangle c dt_c,
\end{equation}
where $t_c$ is comoving time, $\lambda=1/(\sigma_T n_c)$ is the comoving mean free path of photons, $n_c$ is comoving
density, $t_i$ is the initial comoving time, $t_f$ is final comoving time when the photon leaves the outflow. The integral (\ref{int}) should be taken along the average photon path. In the case of optically thick medium this path is given by the
equation of motion of the outflow
\begin{equation}\label{avphpath}
r=r_i+\beta ct,
\end{equation}
where $r_i$ is the laboratory radial position of the photon at initial laboratory time.

In the photon thick case the photons stay in the outflow long after 
decoupling, so $t_f\rightarrow\infty$. Then, taking into account relations the
$n_c=n_l/\Gamma$, $dr=\beta c dt$, and $t_c=t/\Gamma$ along the world-line (\ref{avphpath}), we
obtain
\begin{equation}
 \langle N \rangle=\int_{t_i}^{\infty} \sigma \frac{n}{\Gamma} \frac{c 
dt}{\Gamma}= 2\tau_i,
\end{equation}
where $\tau_i$ is optical depth of the outflow at $r_i$. This result is in agreement with \cite{vurm11}.

In photon thin case the photons leave the outflow from its boundaries, and the time
interval needed for the photon to reach them by random walk can be
estimated as $t^c_{D}=l_{c}^{2}/D_c$, where $l_{c}=\Gamma l$ is the comoving
radial thickness of the outflow, and diffusion coefficient is 
$D_{c}=(c\lambda_{c})/3=c/(3\sigma n_{c})$. When this time is much less than the 
characteristic time of expansion, equal to $t_i$, that is the case when initial 
radius $r_i$ is much larger than radius of diffusion $R_D$, we have
\begin{equation}
 \langle N \rangle\simeq 3\tau_i^2,
\end{equation}

In the opposite case when initial radius $r_i$ is much smaller than 
the radius of diffusion $R_D$, we have
\begin{equation}
 \langle N \rangle\simeq \frac{1}{\Gamma^2}\sqrt{\tau_i\tau_0\frac{R_0}{l}}.
\end{equation}

\end{document}